\documentclass{elsarticle}
\pdfoutput=1  %ARXIV
\usepackage{hyperref}    %ARXIV
\biboptions{sort&compress}

\usepackage[final]{pdfpages}  %ARXIV

\def\pDOSY{\textit{p}--DOSY}

\usepackage[final]{changes}

\begin{document}

    \begin{frontmatter}
    \date{5 June 2015}
    \title{Accurate DOSY measure \replaced{for}{to} out-of-equilibrium systems  \replaced{using}{by} permutated DOSY (\pDOSY)}

    %% or include affiliations in footnotes:
    \author[Wageningen]{Maria Oikonomou\fnref{first_author}}

    \author[IGBMC,NMRTEC]{Julia Asencio Hern\'andez\fnref{first_author}}
    \fntext[first_author]{These two authors contributed equally}
    \author[Wageningen,IRICA]{Aldrik H. Velders}
    \author[IGBMC]{Marc-Andr\'e Delsuc}
    \cortext[mycorrespondingauthor]{Corresponding author}
    \ead{madelsuc@unistra.fr}

    \address[Wageningen]{Laboratory of BioNanoTechnology, Wageningen University, PO BOX 8038, 6700 EK Wageningen, The Netherlands}
    \address[IGBMC]{Institut de G\'en\'etique et de Biologie Mol\'eculaire et Cellulaire, INSERM, U596; CNRS, UMR 7104; Universit\'e de Strasbourg, 67404 Illkirch-Graffenstaden, France}
    \address[NMRTEC]{NMRTEC, bioparc, 1 Bd Brandt, Illkirch-Graffenstaden 67400, France}
    \address[IRICA]{Instituto Regional de Investigaci\'on Cient\'ifica Aplicada (IRICA), Universidad de Castilla-La Mancha, Avda. Camilo Jos\'e Cela, s/n, 13071, Ciudad Real, Spain}

    \begin{abstract}
    NMR spectroscopy is a \replaced{excellent}{perfect} tool for monitoring in-situ chemical reactions.
    In particular, DOSY measurement is well suited to characterize transient species by the determination of their sizes.
    However, here we bring to light a difficulty in the DOSY experiments performed \replaced{in}{on} out-of-equilibrium systems. 
    On such a system, the evolution of the concentration of species interferes with the measurement process, and creates a bias on the diffusion coefficient determination that may lead to erroneous interpretations.

    We show that a random permutation of the series of gradient strengths used during the DOSY experiment allows to average out this bias.
    This approach, that we name \pDOSY\ does not require changes in the the pulse sequences nor in the processing software, and restores completely the full accuracy of the measure.
    This technique is demonstrated on the monitoring of the anomerization reaction of $\alpha$- to $\beta$-glucose.

    \end{abstract}


    \end{frontmatter}

\section*{Introduction}

Diffusion Ordered SpectroscopY (DOSY) is an NMR technique separating components in a mixture spectroscopically according to their translational diffusion\cite{johnson1999, Antalek:2002ww}.
Considered as an NMR chromatographic technique, DOSY NMR has been widely used in life sciences and chemistry for the analysis of
complex mixtures, such as
biological extracts\cite{Barjat:1995kq},
environmental fluids\cite{Morris:1999ig},
supramolecular chemistry\cite{Allouche:2006de,Giuseppone:2008km},
\added{nanoparticles}\cite{VanLokeren:2007,Gomez:2009kf},
or pharmaceutical \replaced{formulations}{analysis}\cite{Gilard:2011wm}.
For spherical shaped components the diffusion coefficient $D$ is inversely proportional to the hydrodynamic radius $r_H$ as described by the Stokes-Einstein equation:
\begin{equation}
D=\frac{k_B T}{6\pi \eta  r_H}
\label{eq1}
\end{equation}
where $k_B$ is the Boltzmann's constant, $T$ is the absolute temperature, and $\eta$ is the viscosity.
Other shapes can be handled either with extended analytical models\cite{Evans:2013bp} or by a statistical description\cite{Auge:2009fv, HernandezSantiago:2015dk}.

A DOSY experiment consists in a series of spin echo or stimulated echo spectra recorded with increasing gradient field strength.
When measured with the Stimulated Echo pulse sequence (STE), diffusion coefficients can be determined by analyzing the signal amplitude as a function of the square of the gradient pulse area using the Stejskal-Tanner equation\cite{Stejskal:1965kr} \added{given here for squared pulses }:
\begin{equation}
I=I_o e^{-D \gamma^2 g^2 \delta^2 (\Delta-\frac{\delta}{3})}                                                                                                 
\label{eq2}
\end{equation}
where $I_o$ is the signal intensity obtained in the absence of field gradient and $I$ in the presence of a gradient of strength $g$, $D$ is the diffusion coefficient, $\delta$ is the gradient pulse duration, $\Delta$ is the effective diffusion delay \added{defined as the time between the start of the bipolar gradient pairs}, and $\gamma$ is the gyromagnetic ratio of the spin under study; \added{see \cite{Sinnaeve:2012jr} for details}

The analysis of the diffusion coefficients is a powerful tool to gain insight on molecular sizes and shapes\cite{Wilkins1999,Gomez:2009kf,Floquet:2009ex},
and can also be used to investigate molecular interactions\cite{Lucas:2004jg, Cohen:2012cx, Nonappa:2015gb}, %, vanDongen:2014fc}
polymer polydispersity\cite{Stchedroff:2004vz, Vieville:2011fj},
or to characterize reactive intermediates\cite{Schlorer:2002ul, Li:2009hl, Nguyen:2009bx, Alvarez:2011tb}.

However, when measuring DOSY experiment on an evolving system, a difficulty arises.
According to the classical Stejskal-Tanner equation, the intensity at each gradient strength depends on the intensity $I_o$ with no gradient applied.
Usually, $I_o$ is measured at the beginning of the DOSY experiment and considered constant throughout the measurement.
However, in out-of-equilibrium systems, $I_o$ is not constant anymore but varies continuously, due to evolution of the concentrations.
%These variations cause the intensity $I$ to depend not only on the applied gradient but also on a varying $I_o$.
Consequently, throughout the DOSY measurement, the signal intensity measured for a given species will depend on the concentration evolution as well as the increasing gradient strength, and will appear to decay either too rapidly or too slowly, depending on the evolution of the concentration, decreasing or increasing over the time of the measurement, creating a bias in the analysis.

The kinetic rates of the out-of-equilibrium system has to be taken into account and be compared to the timeframe of the DOSY measurement.
Rapid DOSY measurements monitoring systems with slow kinetic rates are not affected to a significant degree, because only minor concentration changes occur over the experimental time.
On the other hand, long DOSY measurements monitoring systems with fast kinetic rates are not affected either, but fail to record dynamic information of the system such as reactive intermediates. 

The effect is \replaced{most obvious}{dramatic} for systems with intermediate kinetic rates monitored by DOSY measurements performed in a time similar to the kinetic constant of the chemical systems.
In this case, the DOSY spectra display a \deleted{apparent} shift of the peaks along the diffusion coefficient axis leading to wrong results.

Nilsson et al.\cite{Nilsson:2009ep} proposed the use of trilinear multivariate analysis in order to characterize both the reaction kinetics and the diffusion coefficient, in the analysis of a set of diffusion experiments measured during the chemical reaction.
In this case the concentration variations is used as a third dimension in the multiway analysis.
However in this approach, the DOSY measurements are supposed to be rapid compared to the kinetic rates, and there is no compensation for the unavoidable bias in the diffusion coefficient measure.
To our knowledge, the DOSY studies performed so far on out-of-equilibrium systems\cite{Schlorer:2002ul, Li:2009hl, Nguyen:2009bx, Alvarez:2011tb, Nilsson:2009ep} have not taken into account the analytical bias caused by concentration variations.

We propose here to measure, within a single DOSY measurement, a series of stimulated echo spectra with pulse gradient strengths sampled in a permutated manner, as it was already proposed in order to reduce $F_1$ artifacts in 2D NMR spectroscopy\cite{Bowyer:1999dc}.

\added{A permutation of the list of gradient strengths allows the separation of the experimental time, during which the system evolves, from the gradient strength sampling, thus removing the experimental bias.
Knowing the actual evolution of the concentration, an optimal permutation could eventually be designed.
However this knowledge is usually not available, and such an optimal permutation not accessible, so we choose to perform random permutations of the gradient list.
This permutation distributes randomly the concentration variation over the whole range of gradients, actually transforming an experimental bias into an additional random noise (see figure S14 in Supp. Mat.).
The analysis of the permutated experiment leads to a less precise measure (larger error-bars) but eventually more accurate (less systematic error).
Each realization of the random permutation may remove only partially the systematic error, and the accuracy is improved with a larger number of gradient levels.
Simulations performed for several model systems (Figures S13 to S27 in Supp.Mat.) show that large bias may be observed for a regular approach and are fully reduced by this procedure.
Additionally, the improvement of the measure with larger gradient lists is clearly observed.
}

\begin{figure}
\includegraphics[scale=0.5]{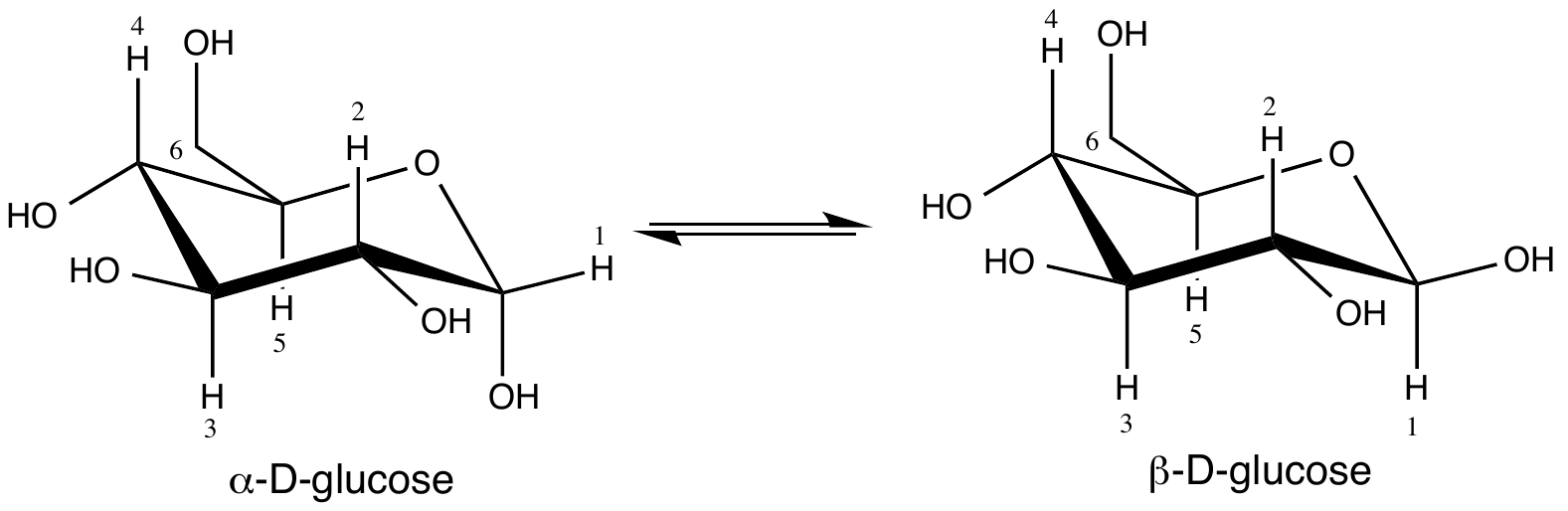}
\caption{Anomerization of $\alpha$-Glucose to $\beta$-glucose with proton numbering}
\label{fig1}
\end{figure}

From the sequential list of pulse field gradient strengths, a random permutation is computed, and the DOSY is measured with a new list of gradient strengths in permutated order.
This protocol has been \replaced{named}{called} permutated--DOSY (\pDOSY).
To test the effect of the permutated acquisition on the accuracy of the \replaced{measurement}{measure},
\added{a set of simulations was performed and}
the diffusion coefficient of glucose was monitored during the equilibration from pure $\alpha$-glucose to the \replaced{anomeric}{anomer} equilibrium of $\alpha$ and $\beta$-glucose\cite{Gurst:1991jv} (see Figure \ref{fig1}).

\section*{Materials and Methods}
\added{Simulations were performed in Python 2.7, using the libraries numpy 1.9 and scipy 0.15.
They are presented in supplementary material, and can be downloaded at 
\url{https://github.com/delsuc/p-DOSY}.}

Crystalized $\alpha$-D-glucose (anhydrous, 96\% purity) was purchased to Sigma-Aldrich (USA).
Deuterated solvents DMSO-d6 (99.96\%) and D$_2$O (99.9\%) were purchased to Euriso-Top (France). 
All samples were measured in 5.0\,mm tubes.
A stock solution of 100.0\,$mM$ $\alpha$-D-glucose in DMSO was prepared.
100\,$\mu$L of the glucose stock solution were added in 900\,$\mu$L D$_2$O resulting in final concentration of 10.0\,$mM$. 

DOSY NMR measurements were carried out both on a Bruker and on a Varian instrument, (for Varian experiments see supplementary material).
The Bruker instrument was a 500 Avance I NMR spectrometer, operating at 500.137\,MHz for $^1$H, equipped with a 5\,mm TXI probe.
The pulse sequence used was a longitudinal eddy current delay bipolar gradient pulse (\texttt{ledbpgp2s})
An exponential gradient list of 32 values was created by using the standard AU program \texttt{dosy}.
For the \pDOSY\ experiment, the list was permutated by an ad-hoc python script (see supplementary material), and the permutated list was used for acquisition and processing.
Experiments were acquired with 8 scans, $\delta/2$ of 3.0\,ms and $\Delta$ of 150\,ms, for an overall time of 23:09\,min.
A series of 15 spectra was acquired at a temperature of 300\,K for a total duration of 6.0\,hours.
Before and after every DOSY spectrum, an 8 scans proton spectrum was recorded.
The whole experiment was duplicated once with regularly increasing series of gradients, and once with a permutated series. 

The DOSY spectra acquired on Bruker were processed with the NMRnotebook program (version 2.7, NMRTEC France), using the DOSY MaxEnt imple-mentation\cite{Delsuc:1998uy,VanLokeren:2008gk}.
The experiments performed on Varian were analyzed with the vnmrj (version 6.3) software.
In both cases, no modification had to be made to the computation protocol despite the permutated entries in the DOSY file.

The final diffusion coefficients were calculated from the average position of the anomeric signal for each species.
Error bars were estimated as the standard deviation along the diffusion axis of the different signals.
%by averaging the values measured for all major peaks in each proton signal.
Minor viscosity variations were compensated by normalizing all measurements to the DMSO diffusion coefficient.

\section*{Results and discussion}

The \pDOSY\ approach, and its capacity to restore accuracy in the determination of diffusion coefficients was tested using glucose anomerization as a test case for out-of-equilibrium systems.
Glucose crystalizes in the $\alpha$ form, as a consequence, this form is dominant in the solution just after \replaced{dissolution}{solubilization}.
The anomerization of $\alpha$ to $\beta$-glucose was followed by acquiring DOSY measurements for approximately 6.0 hours, with $\alpha$-D-glucose varying from  96.6\% to 44.6\%.
The same experiment was performed twice, the first time with the standard DOSY method, and the second time using the \pDOSY\ approach.

\begin{figure*}
\includegraphics[scale=0.5]{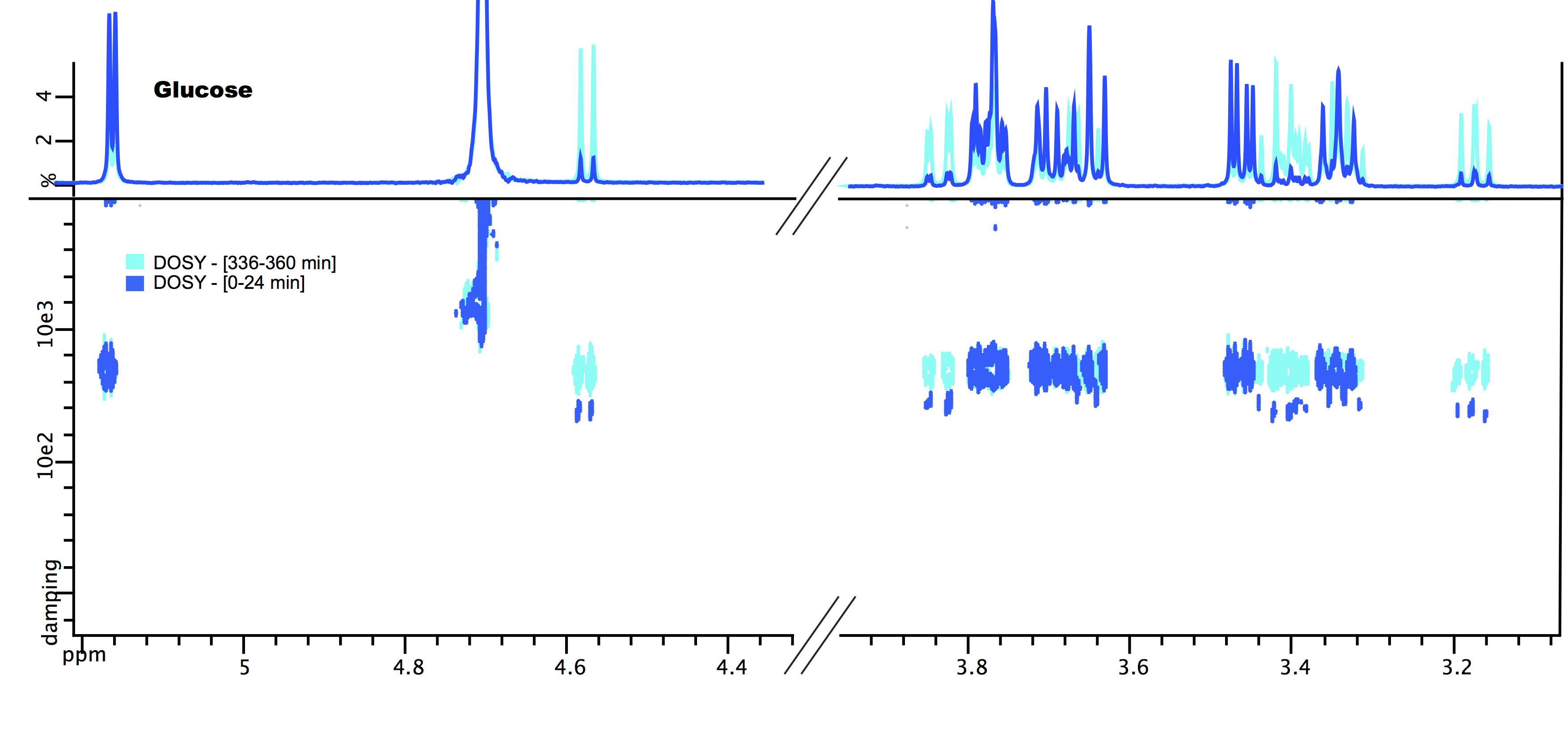}
\caption{Overlay of two standard DOSY  experiments, acquired between: a) out-of-equilibrium (0.00-25.00 minutes) (blue color) and b) at equilibrium (5.41-6.05 hours) (light blue color)}
\label{fig2}
\end{figure*}

The diffusion coefficients in water of both species $\alpha$ and $\beta$-glucose are expected to be very similar, as they have very similar shapes.
However, they appear with quite different diffusion coefficients when measured with a standard DOSY experiment, acquired in a sequential manner, as can be seen in figures \ref{fig2} and \ref{fig4}.
In particular, the signals from the $\beta$-glucose are found initially at a \replaced{markedly}{marked} slower diffusion coefficient and recover the normal value over the course of the experiment.
This is the result of the progressive increase of concentration of $\beta$-glucose over the duration of a single DOSY measurement, which creates a bias in the exponential analysis of the signal.
At the end of the 6 hours experiment, the recovery is not complete.

\begin{figure*}
\includegraphics[scale=0.5]{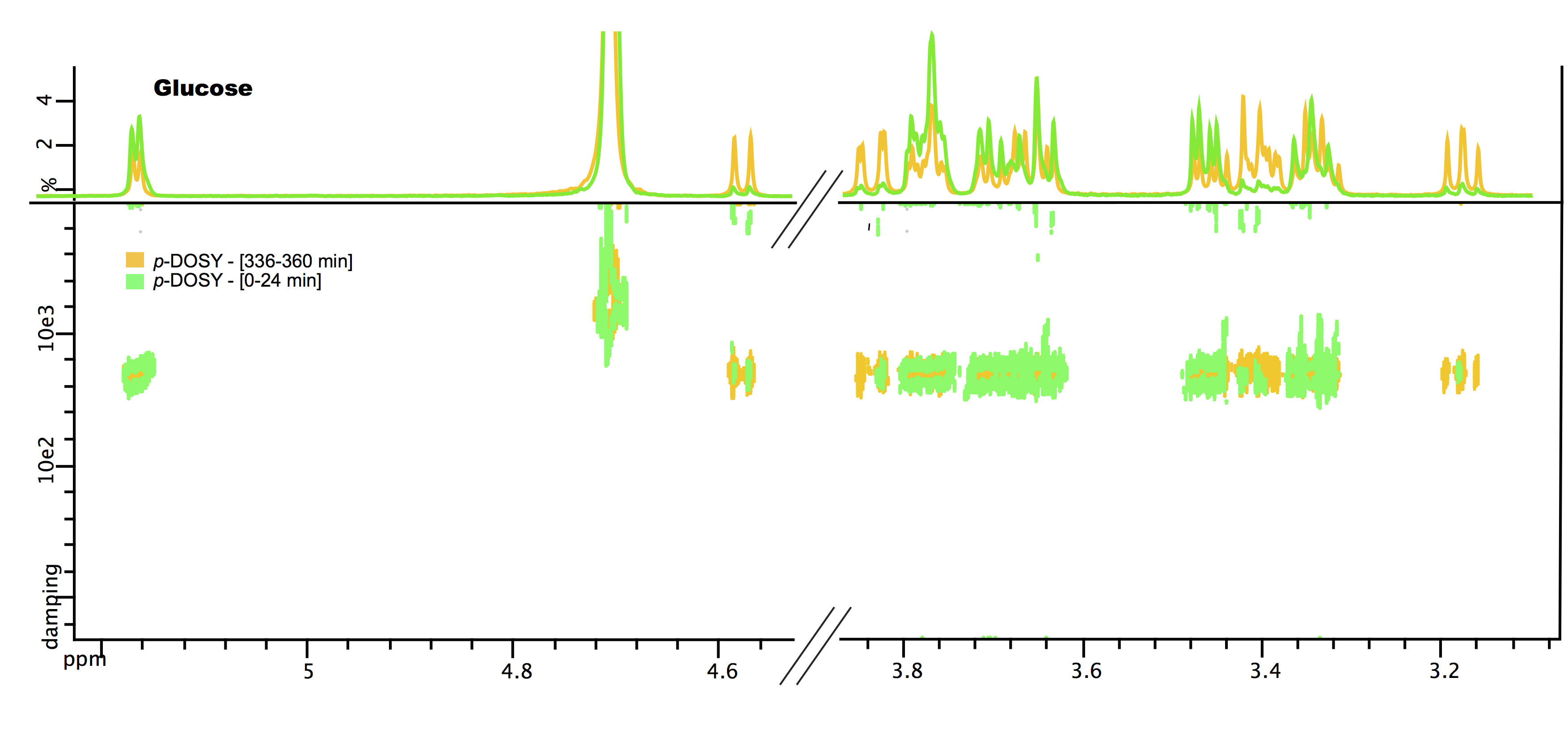}
\caption{Overlay of two \pDOSY\ experiments, acquired between: a) out-of-equilibrium (0.00-25.00 minutes) (green color) and b) at equilibrium (5.37-6.00 hours) (yellow color)}
\label{fig3}
\end{figure*}

\begin{figure}
a)\\
\includegraphics[scale=0.6]{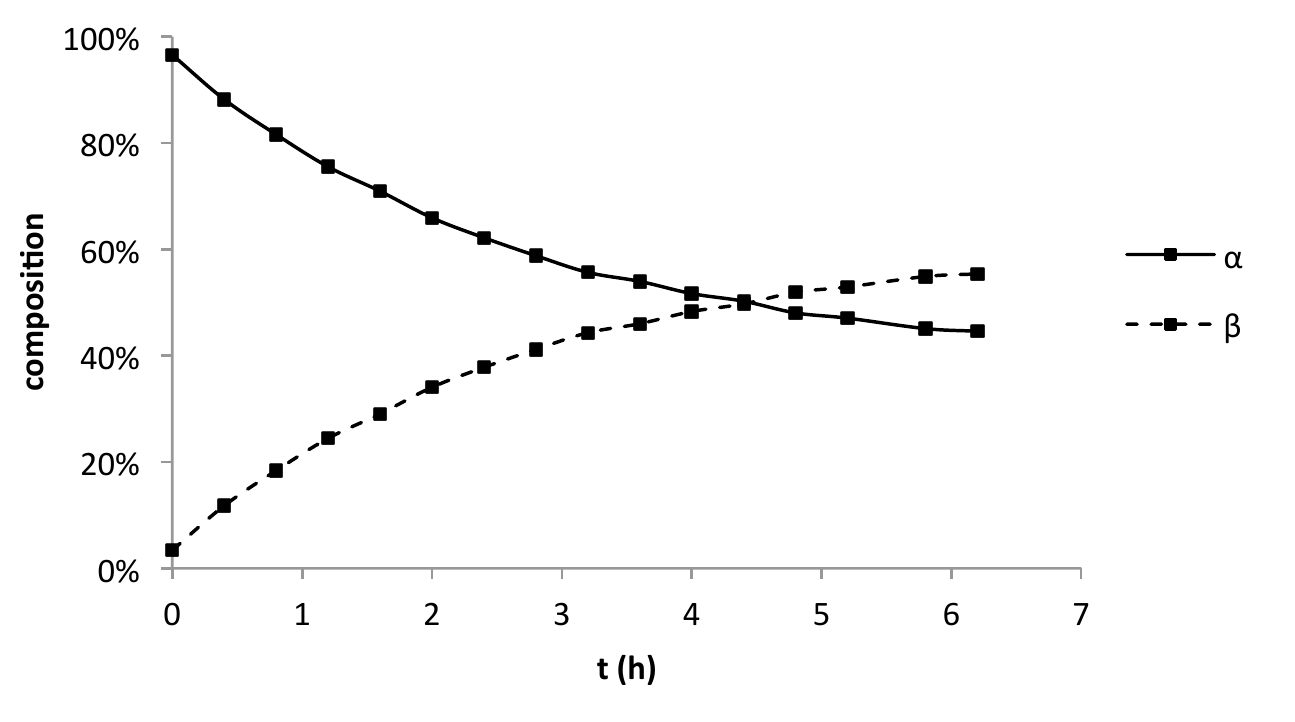}\\
b)\\
\includegraphics[scale=0.6]{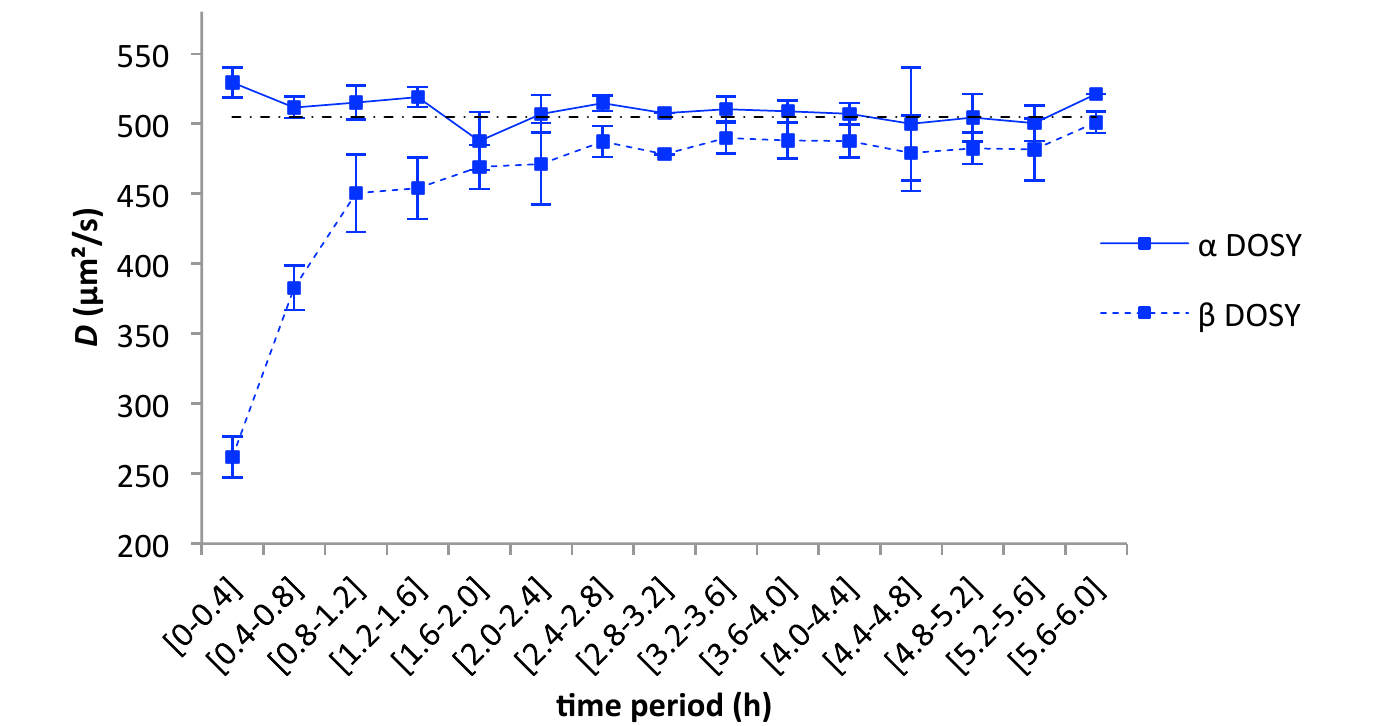}\\
c)\\
\includegraphics[scale=0.6]{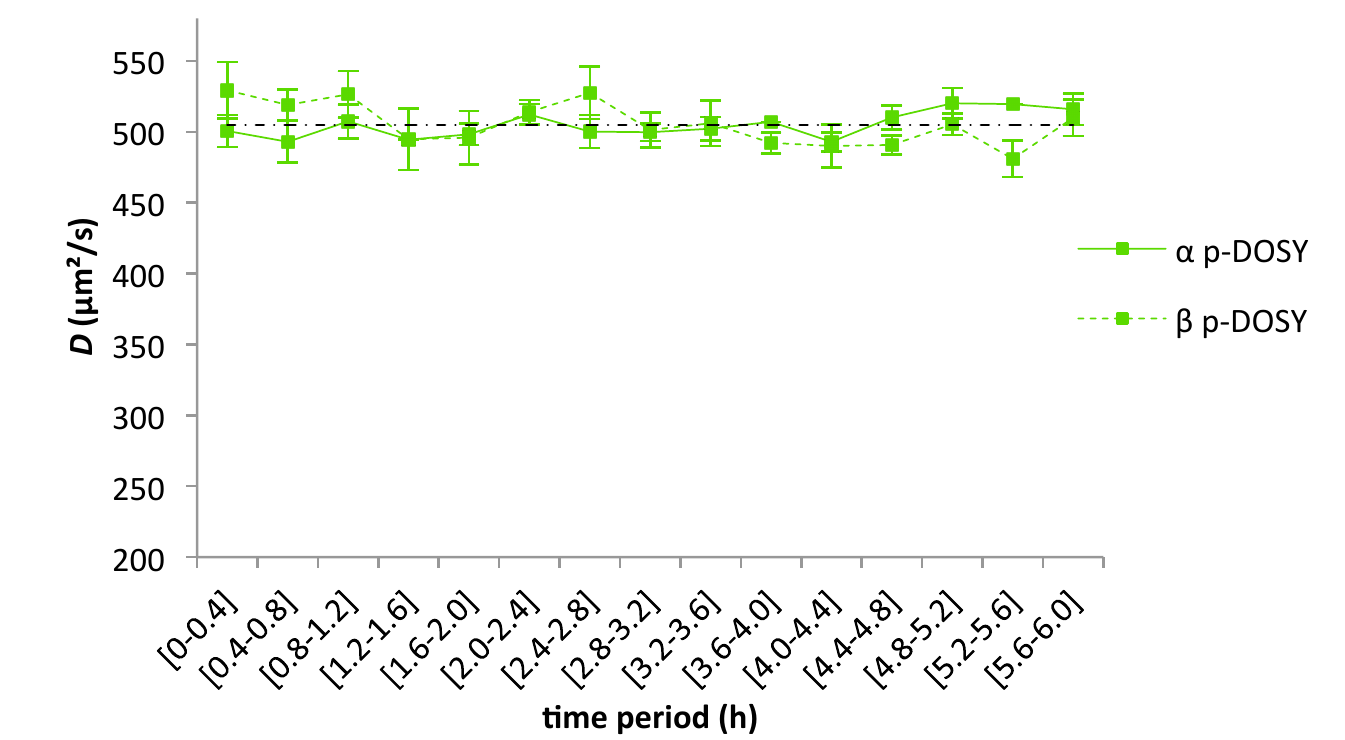}\\
\caption{
a) concentration evolution of $\alpha , \beta$-glucose over experimental time, measured as the intensity of the anomeric proton peak based on 1D experiments interleaved during the \pDOSY\ experiment series;
b) evolution of the apparent diffusion coefficients of $\alpha$ and $\beta$-glucose over experimental time based on a series of sequential conventional DOSY measurements;
c) same as b, obtained with the \pDOSY\ experiment.
In b) and c) the horizontal doted line is located at the diffusion coefficient of  $\alpha$-glucose, 
measured as the mean value of \pDOSY\ value over the whole \replaced{period}{kinetics}. }
\label{fig4}
\end{figure}

On the other hand, this bias is completely absent when the anomerization equilibrium is monitored using the \pDOSY\ measurements, run \replaced{under}{in} the same conditions (see figure \ref{fig3}).
 
In both experiments, the diffusion coefficients of water and DMSO display the standard values for all experiments run during the kinetics, indicating that the error on $\beta$-glucose is not due to an experimental bias but \replaced{solely due}{uniquely} to the variation of the concentration.
Figure \ref{fig4} presents the evolution of the DOSY signal intensities and of the apparent diffusion coefficients over the course of the experiment.
It shows that the initial $\beta$-glucose concentration is very low, and that the two populations equilibrate slowly, so that 6 hours are not sufficient to reach the complete equilibrium.
When measured with the standard DOSY experiment, the apparent $\beta$-glucose diffusion coefficient, varies in an anomalous manner before converging toward the correct value, while the \pDOSY\ measure presents stable measurements for both species.

The poor accuracy of the diffusion coefficient of $\beta$-glucose is attributed to miscalculation of the analysis of the reference intensity, $I_o$. 
The diffusion coefficient of $\beta$-glucose consequently, appears to have an incorrect low value.
This effect is stronger in the first DOSY measurements since there is a higher kinetic rate.
Similarly, the $\alpha$-glucose that evolves in the opposite direction, displays an apparent diffusion coefficient slightly larger than expected in the first DOSY measurements.
The effect is less marked because the relative concentration variation of the $\alpha$-glucose is less important than in the case of $\beta$-glucose.

On the other hand, the experiment performed with a series of \pDOSY\ measurements does not display this systematic error, and accurate values are determined \replaced{throughout}{over} the whole experiment.
Because of the dispersion effect of the random permutation operation, the systematic evolution of the concentration is  transformed here into random errors on the $I_o$ value.
This added random noise has no effect on the analysis, but disperses in a random manner the concentration variation over the gradient list, thus providing a way to average out bias.
However, because of the added noise, a slightly less precise determination of the diffusion coefficients is expected.
This effect, while not very strong, can be observed on Figure \ref{fig4} by the evolution of the error bars on the \pDOSY\ determined diffusion coefficients.
The error bars present a somewhat larger spreading in the beginning of the kinetic and converge toward a minimal value.

In a nutshell, one can say that the \pDOSY\ experiment provides on evolving chemical systems, a slightly less precise, but much more accurate determination of the diffusion coefficient that the standard DOSY experiment.

\section*{Conclusion}
The accurate estimation of diffusion coefficients is challenging in systems away from equilibrium. 
An inherent characteristic of out-of-equilibrium systems is the evolution of the concentration of the different components during a DOSY NMR experiment. 
Since the basis of a single DOSY measurement is the analysis of \added{the} exponential decay of the signal with increasing gradient strength. Concentration evolution, within the characteristic measurement time, creates a bias, which can lead to large errors in the diffusion coefficient estimation. 
This was observed for the test case of the anomerization of glucose.
Randomizing the list of gradient strength arrays within a single DOSY measurement, decouples the gradient evolution from the concentration evolution and removes the bias in the analysis, thus restoring the accuracy of the measure. 
The concentration evolution is still there though, as a noise source, and may \deleted{have} impact the precision of the measure compared to a static sample, although its impact is \replaced{minimal}{minimum}. 
Depending on the spectrometer software, this experiment requires little or no modification of the standard procedure, and the produced data-set can easily be analyzed using regular processing programs.

Altogether, the DOSY experiment with a randomized list of gradients (the permutated DOSY or \pDOSY ) is an \replaced{unbiased experiment in the}{experiment exempt of bias in} presence of an evolving system, in contrast to conventional DOSY.
It is perfectly suited for the analysis of chemical reactions and out-of-equilibrium systems and can be used to monitor the rate of reaction kinetics,
to characterize \emph{in-operando} transient species and measure their diffusion coefficient\added{s}, 
or to observed molecular organization phenomena and dynamics effects.

The robustness of \pDOSY\ can also be used to protect diffusion results from any experimental artifact, such as temperature shift or spectrometer drift
\added{when such artifacts are known and cannot be easily compensated for.}
As it does not add any burden in the acquisition step nor at the processing step, we recommend spectroscopists to use \pDOSY\ \replaced{in this experimental cases.}{used systematically in diffusion and DOSY experiments.}

\section*{\added{Supplementary Information}}
\added{Supplementary Information is available with this work.
Figures S1-S2 present additional glucose spectra.
Figures S3-S6 compare DOSY and \pDOSY\ measured on a Varian spectrometer.
Figure S7 presents the python program used for computing permutations.
Figures S8-S27 present the results of the various simulations, the siumlation program can be downloaded at \url{https://github.com/delsuc/p-DOSY}.}

\section*{Acknowledgments}
JAH was supported by the \replaced{Marie Curie}{european} ITN ``ReAd'' Network.
MO and AHV thank the Marie Curie ITN ``Dynamol'' for support.
\added{The authors acknowledge reviewer \#1 for constructive remarks on the effect of the number of the gradient levels.}
Part of this work was supported by Instruct from the European Strategy Forum on Research Infrastructures (ESFRI) and supported by national member subscriptions, the French Infrastructure for Integrated Structural Biology (FRISBI) (ANR-10-INSB-05-01)

\section*{References}

\bibliography{pDOSY}



\section*{Supplementary material}

# Supplementary Material

# Accurate DOSY measure of out-of-equilibrium systems by permutated DOSY (p–DOSY)

Maria Oikonomou[a,1], Julia Asencio Hernández[b,c,1], Aldrik H. Velders[a,d], Marc-André Delsuc[b]

[a] Laboratory of BioNanoTechnology, Wageningen University, PO BOX 8038, 6700 EK Wageningen, The Netherlands
[b] Institut de Génétique et de Biologie Moléculaire et Cellulaire, INSERM, U596; CNRS, UMR 7104; Université de Strasbourg, 67404 Illkirch-Graffenstaden, France
[c] NMRTEC, bioparc, 1 Bd Brandt, Illkirch-Graffenstaden 67400, France
[d] Instituto Regional de Investigación Científica Aplicada (IRICA), Universidad de Castilla-La Mancha, Avda. Camilo José Cela, s/n, 13071, Ciudad Real, Spain
1 These two authors contributed equally

1. **Glucose study by NMR**

The *α,β-D-glucose* was studied by NMR and the assignment of the molecule have been done, to be able to characterize the a *α* and *β* protons for further analysis. This assignment is shown in the 1D spectrum in Figure S1 and it corresponds to the numbering of both anomeric molecules in Figure 1.

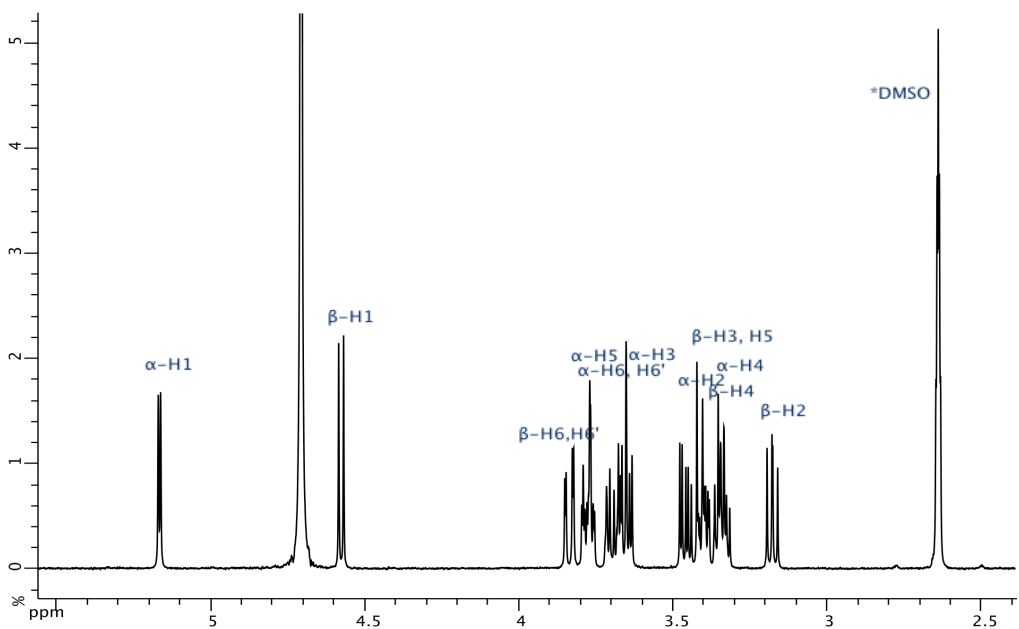

*Figure S1: Proton assignment of a mixture of α,β-D-glucose (42.7% α-D-glucose and 57.4% of β-D-glucose) at equilibrium*

## 2. *p*-DOSY

The DOSY experiment and the *p*-DOSY experiment, measured on α-glucose sample after solubilisation at the beginning of the anomerization kinetics are shown superimposed in Figure S2. The bias on the diffusion coefficient can be seen for the $β\text{-}H_1$.

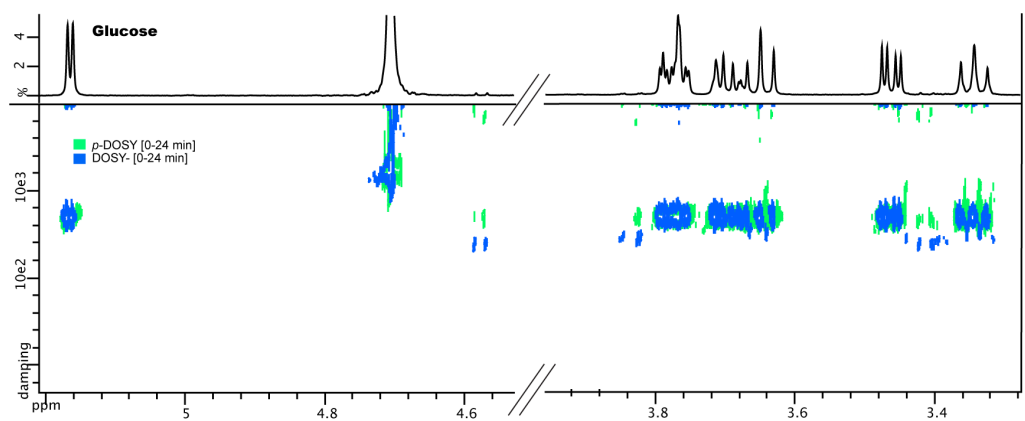

*Figure S2: Overlay of two DOSY measurements, acquired between 0.0-25.0 minutes, by p-DOSY arrays (green color) and sequential (blue color) DOSY arrays*

## 3. Varian experiments

The Varian instrument was an Inova 500 NMR spectrometer operating at 499.7722 MHz for $^1H$, equipped with a four-nucleus 5mm $^1H(^{15}N\text{-}^{31}P)$ PFG High-Field Indirect detection probe. A DOSY bipolar pulse pair stimulated echo with convection compensation (`Dbppste_cc`) was used.
Except for a temperature of 298.0 K, the experimental set-up equivalent to the Bruker experiments. For the *p*-DOSY experiment, the permutated pulse gradient strength arrays was obtained by activating the randomization flag.
DOSY experiments were run in four different modes: sequential, interleaved, permutated, and permutated and interleaved combined.
The sequential DOSY spectra were acquired with 32 gradient increments with increasing gradient strengths. In the interleaved experiment, the block size was set to 4 (4 sets of 32 increments of 4 scans each). For the permutated experiment, 32 increments recorded by permutating the gradient strength pulses. A combination of both interleaved and permutated was also acquired (4 sets of 32 increments of 4 scans each with permutated gradient pulses).

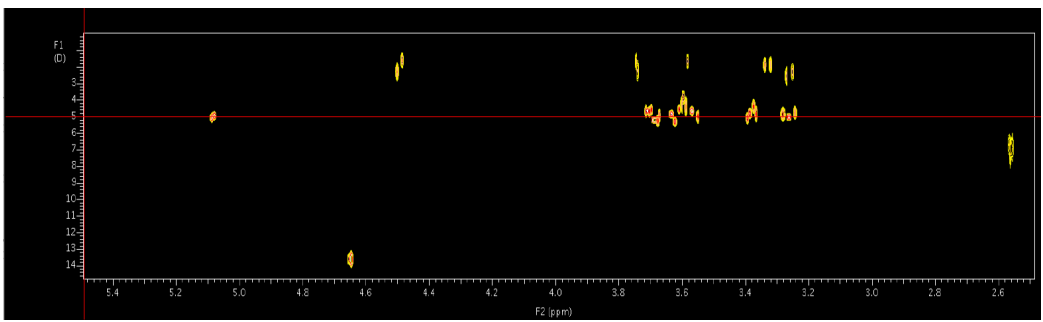

*Figure S3: Sequential DOSY measurement acquired between 0. 0-25.0 minutes (13.0-50.0% β-D-glucose) in Varian*

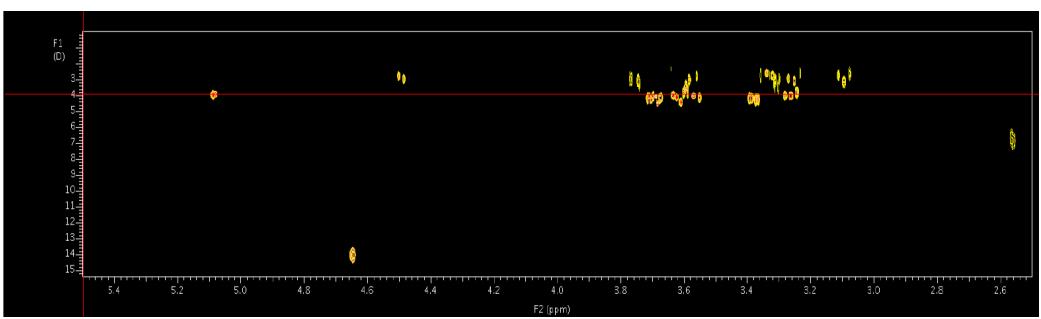

*Figure S4: Interleaved DOSY measurement acquired between 0.0-25.0 minutes (13.4-42.5 % β-D-glucose)in Varian*

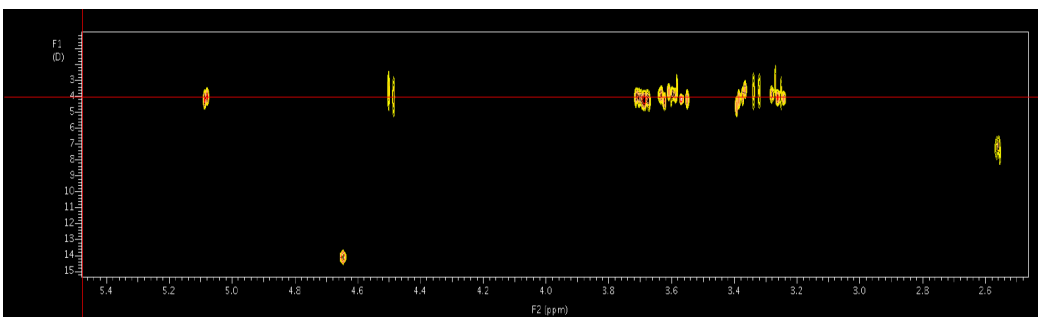

*Figure S5: P-DOSY measurement acquired between 0.0-25.0 minutes (8.3-40.8% β-D-glucose) in Varian*

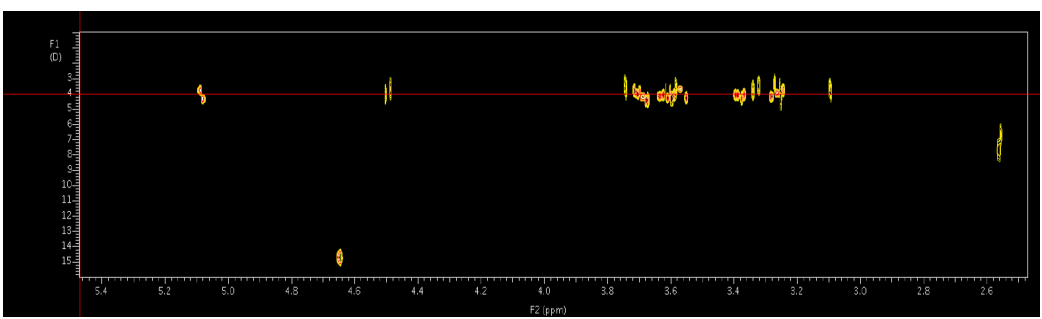

*Figure S6: Combination of permutated and interleaved DOSY measurement between 0.0-2.0 (7.34-33.77% β-D-glucose) in Varian*

### 4. Permutated *Difframp* and *difflist*

Here is presented the ¨*Permute*¨ script to permutate the Difframp and difflist for acquisition and processing the *p*-DOSY experiments.

The original *Difframp*, created by TopSpin when a DOSY experiment is launched, has to be replaced by the permutated *Difframp* before running the *p*-DOSY experiment. The same procedure has to been followed for the *difflist* before processing the spectrum.

```python
N = 64    # N is the length of the gradient list
          # Adapt to your experiment
perm = random.sample(range(N), N)

print N, "points :", perm
filelist = ['difflist', 'Difframp'] # difflist is for processing, Difframp is for acquisition
for fname in filelist:
        F = open(fname,'r')
        FF = open(fname+'P','w')
        ll = []
        for l in F:
           if not l.startswith('#'):
               ll.append(l)
           elif l != '##END=':
               FF.write(l)
        F.close()
        if len(ll) != N:
            raise Exception( 'This prgm is meant for %d points'%N )
        for i in range(N):
            FF.write(ll[perm[i]])
        FF.write('##END=')
        FF.close()
```

python `permute.py` script used to compute a random permutation of the initial your Difframp and difflist files, for acquisition and processing of the p-DOSY experiment.

    %ARXIV

\end{document}